# Why is FPGA-GPU Heterogeneity the Best Option for Embedded Deep Neural Networks?

Walther Carballo-Hernández
Université Clermont Auvergne
CNRS, SIGMA Clermont, Institut Pascal
F-63000 Clermont-Ferrand, France
Email: walther.carballo_hernandez@uca.fr

Maxime Pelcat
IETR, UMR CNRS 6164, Institut Pascal
UMR CNRS 6602, Univ Rennes, INSA
Rennes, France
Email: mpelcat@insa-rennes.fr

François Berry
Université Clermont Auvergne
CNRS, SIGMA Clermont, Institut Pascal
F-63000 Clermont-Ferrand, France
Email: francois.berry@uca.fr

*Abstract*—Graphics Processing Units (GPUs) are currently the dominating programmable architecture for Deep Learning (DL) accelerators. The adoption of Field Programmable Gate Arrays (FPGAs) in DL accelerators is however getting momentum. In this paper, we demonstrate that Direct Hardware Mapping (DHM) of a Convolutional Neural Network (CNN) on an embedded FPGA substantially outperforms a GPU implementation in terms of energy efficiency and execution time. However, DHM is highly resource intensive and cannot fully substitute the GPU when implementing a state-of-the-art CNN. We thus propose a hybrid FPGA-GPU DL acceleration method and demonstrate that heterogeneous acceleration outperforms GPU acceleration even including communication overheads.

Experimental results are conducted on a heterogeneous multi-platform setup embedding an Nvidia® Jetson TX2 CPU-GPU board and an Intel® Cyclone10GX FPGA board. The SqueezeNet, MobileNetv2, and ShuffleNetv2 mobile-oriented CNNs are experimented. We show that heterogeneous FPGA-GPU acceleration outperforms GPU acceleration for classification inference task over MobileNetv2 (12%-30% energy reduction, 4% to 26% latency reduction), SqueezeNet (21%-28% energy reduction, same latency), and ShuffleNetv2 (25% energy reduction, 21% latency reduction).

## I. INTRODUCTION

Internet of Things (IoT) and the emerging adoption of heterogeneous architectures in edge devices are currently extending the possibilities of Deep Learning (DL)-powered applications. Indeed, in order to keep reasonable device energy consumption, embedded platforms have started to adopt heterogeneous architectures to keep up with an ever-growing computational demand. While GPUs currently dominate programmable DL acceleration, state-of-the art is still divided on deciding in which cases an FPGA outperforms a GPU as an efficient DL hardware substrate. The main motivation for heterogeneous solutions is to increase computational efficiency through acceleration for a subset of tasks on a full workflow. However, this gain does not mean that the communication overheads induced by inter-layer transfers can be compensated. In this paper, we propose and evaluate FPGA and GPU DL module implementations separately against a heterogeneous solution. Comparisons are based on widely used CNN building blocks using a throughput-optimised pipe-lined Direct Hardware Mapping (DHM) technique for FPGA CNN kernels deployment [1]. The DHM technique incorporates several differences in comparison with conventional GPU network execution. The first difference is the use of a fixed-point computation approach. This compression technique allows us not only to reduce the memory complexity of features and weights, but also to use specialized hardware dedicated to fixed-point computation. In this study, we use 8-bit fixed-point representation as suggested in [2], to avoid affecting heavily the resulting DL accuracy. Secondly, the number of external memory accesses from the device must be considered. Since DHM is based on a stream processing paradigm while keeping parameters and features close to each other, it widely deviates from the memory hierarchy approach of the GPU memory model.

In the three case studies, we aim to evaluate the inference deployment of embedded CNN models such as MobileNetV2 [3], ShuffleNetV2 [4] and SqueezeNet [5], on an embedded FPGA-GPU heterogeneous platform. Although both hardware architectures have been well studied and evaluated on High Performance Computing (HPC) centers, their specific capabilities are still to be exploited on embedded design. In this work, we compute an energy and latency estimation for multiple layers used in these CNN models. We then propose a heterogeneous version of grouped or depth-wise convolution partitions for layer-fusing when allowed by the network architecture at a module-level.

The contributions of this work consist of:
1) demonstrating that DHM on an FPGA is a viable alternative to GPU deep learning acceleration in terms of energy, latency and throughput. However, DHM is currently limited to small layers, due to its extensive usage of FPGA logic resources.
2) comparing the obtained measurements against an embedded GPU implementation for specific layers and operations at a module-level.
3) demonstrating that a combination of GPU and FPGA effectively outperforms homogeneous solutions, even when inter-systems communication overheads are considered.

## II. RELATED WORK

Heterogeneous computing has been increasingly adopted in the last few decades as a result of the power and memory walls of computing. These programmable computing nodes have a diversity of hardware capabilities, different ways to execute instructions, or multiple operation management methods [6].

In cases where there is enough (data or task) parallelism that can be exploited by scheduling, combining FPGAs and GPUs can offer a significant performance [7]. Recent studies feed the discussion in the context of embedded vision applications, comparing for instance an ARM57 CPU, a TX2 GPU and a ZCU102 FPGA [8]. They prove that FPGAs are a better solution than GPUs for more complex pipelines like image filtering, feature extraction or geometric extraction, which is the case in our study. [9] and [10] are the closest works to this study in terms of hardware architecture and partitioning between GPUs and FPGAs in embedded image processing terms. However, the granularity of the partitions are either too fine to affect the communication bottleneck, or too coarse to fully exploit resource allocation. In this paper, we propose heterogeneous partitioning at a module level on state-of-the-art CNNs and compare quantitative results to [9] and [10] in Table I.

In [11], closer to the current study from a communication perspective, a heterogeneous platform consisting of two programmable logic devices. Both are interconnected to be tested on image processing techniques such as histogram of oriented gradients. While some speed-ups are achieved, the inter-subsystems communication through a Peripheral Component Interconnect Express (PCIe) link tends to reduce speed-ups, resulting in a bottleneck. Adopting a host-guest computing structure, more recent works [12], [13] alleviate this bottleneck by bypassing or skipping data allocation at host memory, keeping data in the guest device for a longer time. Shaping memory transfers is critical in a DL context in order to increase the number of layers or parameters to be mapped on the most efficient accelerator and be sent back to a host, as presented in these papers.

## III. PROBLEM DEFINITION

In this section we describe the performance of individual architectures, i.e. of a full implementation on an FPGA with DHM and on a GPU for Section IV models. Two metrics were considered on each device for this work; processing latency ($LAT$) and energy ($E$). We further develop both solutions in a heterogeneous manner showing the results comparison in V:

### A. DHM for FPGA synthesis definition

In this work, we use a data-driven approach that fully exploits the resources in an FPGA to concurrently map and execute multiple CNN layers on the device as a pipeline. DHM was first introduced in [1] as a technique to map processing nodes to logical elements or Digital Signal Processors (DSPs). The synthesized accelerators using this technique take advantage of the fused layers, further explained in section IV, since the intermediate feature maps are stored internally in the device, as well as the kernel weights. This storage avoids the bottleneck communication of intermediate data external memory accesses, increasing energy, latency and throughput efficiency. Additionally, all weights are stored closer to the logic elements, so no external memory accesses are needed for weight retrieval, which in DL applications introduces a considerable overhead. Although this method offers an indisputable high performance efficiency gain, this comes at the cost of an enormous resource requirement. As a consequence of this constrain, only small designs can be mapped using DHM.

Considering the opportunities and limitations of DHM, its usage for CNN acceleration must be handled carefully. The combination of DHM on a heterogeneous platform with the objective to reduce memory accesses on the GPU proves to be an efficient solution, as it is discussed in V. We show, that in fact, while the FPGA is more efficient than the GPU in all evaluation metrics on small kernels, combining such local FPGA acceleration with global GPU processing leads towards the optimal performance.

### B. High-level CUDA code generation for GPU CNN deployment

The hardware architecture and memory model of Nvidia GPUs are highly specialized for batch processing based on Single-Instruction Multiple-Data (SIMD) parallel execution. This execution method requires a specific coding paradigm that handles memory accesses and scheduling to the memory hierarchy. GPUs embed memories with different access latencies, accessed from computing elements called Compute Unified Device Architecture (CUDA) cores. Therefore, both the latency and energy performances are highly dependant of how the kernels threads are executed and how they use this hierarchy.

For CNN applications, multiple levels of data parallelism and data reuse can be achieved by techniques like loop unrolling, tiling or batching; which directly affect hardware performance. Fortunately, because of the wide adoption of high level compiling tools and open source projects, optimized software such as Pytorch [14] alleviate this task for the developer. In this work, we deploy inference for CNNs using the generated CUDA code on different sizes of convolutional layers. Figure 1 shows an example of obtained measurements metrics for latency (Figure 1a) and energy (Figure 1b). The figures are obtained by measuring the execution of a convolutional layer with an input tensor of dimensions 224x224, 3 input channels, from 2 to 64 kernel filters and different kernel sizes. It can be observed that the FPGA implementation outperforms the GPU solution both in terms of energy and of latency. However, the FPGA with DHM deployment is quickly limited by the number of available resources, constraining the depth of convolution filters that can be directly mapped; 64 filters of size $5 \times 5$ in this case.

## IV. STATE-OF-THE-ART CNN MODULES AND GPU-FPGA PARTITIONING

The main motivation for the deployment of heterogeneous platforms for CNN networks is the presence of parallelism and heterogeneity in both CNN computation and communication. The main and more time-consuming operation of the presented building blocks is the convolution operation. Therefore, in order to be able to accelerate execution, it is essential to fully understand its computing model and relevant parameters.

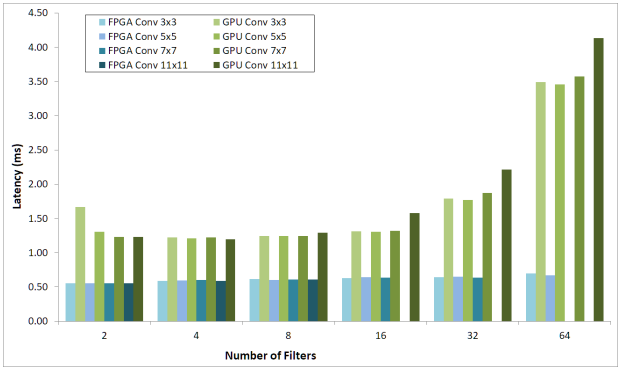 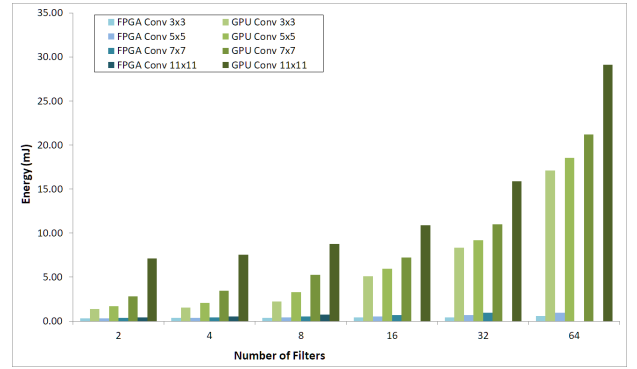

(a)            (b)

Fig. 1: Latency (a) and Energy (b) comparison between multiple convolution function sizes on Cyclone10GX FPGA (blue) and Jetson TX2 GPU (green) for different CNN layers on an input image of 224x224x3. Blue bars represent the layers implemented on the FPGA and the green bars represent the energy consumption on the GPU. The performance factor in this measure is increased result of multiplication on both power and latency metrics.

A convolutional layer (Conv) takes as an input a multidimensional tensor, called Input Feature Map (IFM), $I$ of size $H_I \times W_I \times C_I$ from a previous layer $l-1$ and it is multiplied and accumulated with a sliding window of a kernel tensor $K$ of size $k_h \times k_w \times C_I \times N$. Typically in most applications $k_h = k_w$. The resulting Output Feature Map (OFM) $O$ on the current layer $l$ is obtained from Multiply-and-ACcumulate (MAC) operations and inter-layer communication. Recent CNN algorithmic optimizations are constantly introducing non-regular patterns into the networks in the form of different layer types with a variety of operations. In this subsection, we describe the main building blocks or modules and their partitioning from the current mobile CNN models:

- **Depth-Wise separable Convolution (DWConv):** This technique was first described in [15] and fully utilized in [3]. The main concept relies upon a sort of factorization of a traditional convolutional layer. The first of the resulting operations is a $k \times k$ convolution over every single input channel. The second operation over this tensor is a $1\times 1$ convolution, i.e. a scaling with a depth of $C_I$ resulting in the first channel of the output tensor. Figure 2a shows a layer $d_l$ as a DWConv. We propose a partitioning delegating all the $1\times 1$ convolution on the FPGA for all layers. This allows us to reduce the number of weights directly mapped on the device, saving valuable resources using DHM.
- **Grouped Convolution (GConv):** This partitioning method divides the computational load into workflows that can be executed in parallel and concatenated afterwards. In Figure 2b two contiguous partitions of different sizes are created for each device. The GPU partition takes the subset of the IFMs $H_I \times W_I \times (C_I - g_l)$ and the filter tensor of size $k \times k \times (C_I - g_l) \times N$, while the FPGA takes $H_I \times W_I \times g_l$ and the filter tensor of size $k \times k \times g_l \times N$.
- **Fused-Layer:** It was first introduced in [16] as a method to store intermediate weights and neuron activity in cache from adjacent layers in depth. This approach handles one of the most common challenges in CNN models, the data transfer burden. In Figure 2c the $f_l$ number of parameters of layer $l \in L$ is internally stored on the FPGA to be executed in a pipe-lined fashion [17]. The OFM of the last layer in the partition is then transferred to the GPU.

## V. EXPERIMENTAL METHODOLOGY, EVALUATION AND RESULTS

In this section we describe the experimental methodology deployed to obtain the proposed metrics. In section V-A, we discuss the experimental setup and how individual performance metrics for each device are obtained. In section V-B we present the results of the heterogeneous platform measurements for different operations with the layer-wise partitioning from section IV.

Figure 3 shows the selected embedded computing nodes as case study on a custom prototyping board, linking both devices by a communication node or interface. Additionally, this section describes more in the detail the experimental setup as case of study and baseline for measurement metrics comparison.

### A. Measurement-based energy and latency performance comparison

On the Jetson TX2 Module-on-Chip (MoC), a Tegra TX2 System-on-Chip (SoC) is incorporated, which at the same time, includes an integrated multi-channel power monitor. The ImageNet pre-trained mobile CNN models were obtained from Pytorch [14] and the torchvision model zoo. On the FPGA side, we use the Power Estimation tool® from Intel Quartus Pro Edition® targeting multiple convolutional task operations on the Intel® Cyclone10GX FPGA. The function synthesis is based on the DHM technique described in III. DHM maps directly the function on hardware. Therefore, its power varies rapidly with the number of processing elements and registers mapped on the device. In Figure 1 an example of energy efficiency comparison between both devices is shown. It can

21

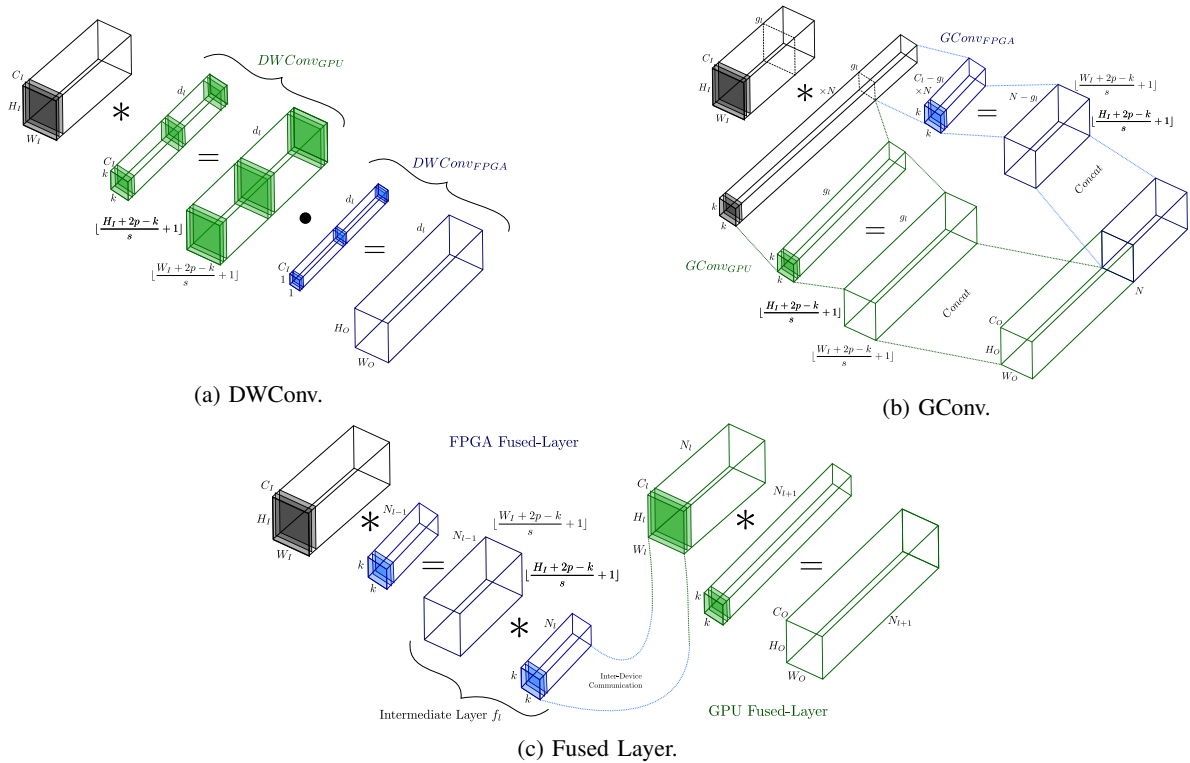

(a) DWConv.

(b) GConv.

(c) Fused Layer.

Fig. 2: Proposed GPU and FPGA layer-wise CNN mappings. The partition in blue represents the data produced on the FPGA, while in green the data produced on the GPU. (a) Depth-wise convolution example where the $k \times k$ convolution is executed on the GPU and the $1\times1$ convolution is executed on the FPGA. (b) Grouped convolution example where the $C_I$ input channels and kernel filters are divided on each device. (c) Fused layer example where a couple or intermediate layer activity are stored in the internal FPGA on-chip memory.

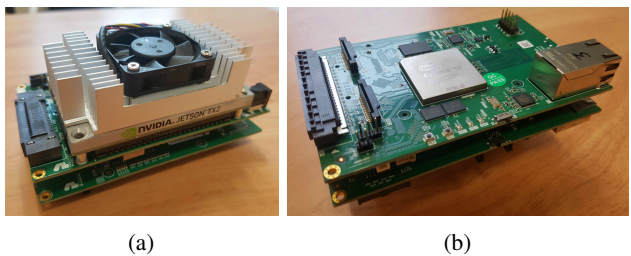

(a)          (b)

Fig. 3: Heterogeneous prototype board consisting of (a) an embedded CPU-GPU Nvidia® Jetson TX2 System-on-Module (SoM) at the top board and (b) an Intel® Cyclone 10GX FPGA at the bottom board interconnected by a 4-lane PCIe gen2 interface.

be observed that the FPGA has a better energy efficiency that outperforms the GPU with orders of magnitude. This effect increases with the number of kernel filters on a fixed IFM. Nonetheless, this is only true as long as the design fits on an embedded FPGA device, like the Cyclone 10 GX FPGA.

### B. Evaluation and results

Given the metric measurements on individual devices and the data-flow graphs for the heterogeneous platform from the proposed partitioning, we validate and evaluate their efficiency on the hardware configurations described from the architecture model in Figure 3. For a fairer comparison between the monolithic homogeneous GPU-only setup and the heterogeneous FPGA-GPU evaluation, both setups were tested with the same configuration parameters and task workloads. The selected CNN models were pre-trained with the ImageNet dataset. Layers hyper-parameters, i.e. IFM and OFM, were obtained from the original papers.

To keep up with a better model precision the first two IFMs dimensions of the layers, $H_I$ and $W_I$, are sampled following the typical architecture tensor sizes of 224x224, 112x112 and so on down to 4x4. This allows us to fully exploit the dimension reduction of the IFM result of the GConv. Because our hardware setup is highly bounded by the PCIe throughput of 2.5GBytes/s, these observations are crucial to keep up with a good performance.

From Figure 4a it can be observed a comparison between the energy in mJ and latency in ms of the layers from SqueezeNet. The heterogeneous solution has a significant energy efficiency gain up to $28\%$ with no significant impact on the latency. This is mostly because the energy efficiency of the $Conv3x3$ task on the FPGA is higher than that on the GPU. In the case of the latency, because both the time spent in communicating

22

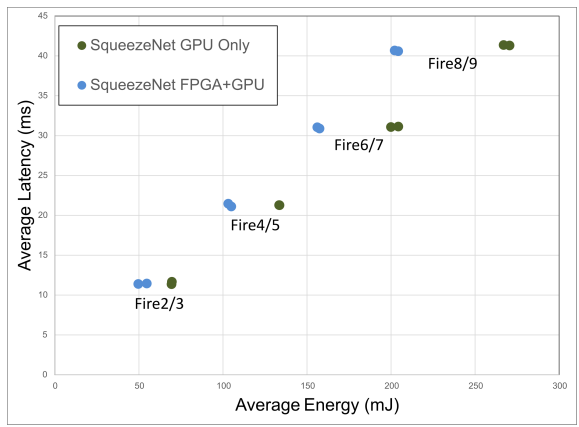
(a) SqueezeNet's layers performance.

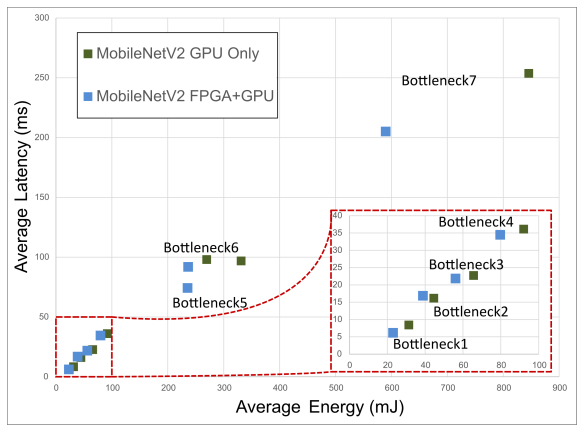
(b) MobileNetv2's layers performance.

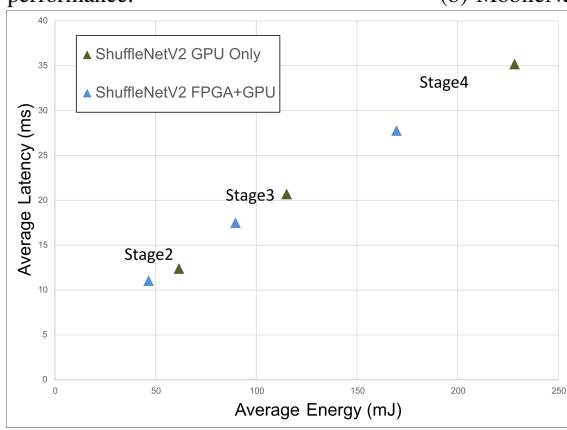
(c) ShuffleNetv2's layers performance.

Fig. 4: Average metric performance space of the tested with different workloads for (a) SqueezeNet's (b) MobileNetv2's and (c) ShuffleNetv2's on an homogeneous GPU-only platform (green) and our FPGA-GPU heterogeneous platform (blue). $x$-axis represents the average energy and $y$-axis the average latency.

between devices and the processing time on the FPGA are still shorter than the execution of the $Conv1x1$ task on the GPU, it is possible to hide its latency during the execution time of the GPU. This means that if the latency of the FPGA and the communication is less than the GPU latency, then the max function as consequence of the heterogeneous model's parallel execution, will be dominated by the GPU-side latency. This is highly beneficial because, this sub-task is small enough, thanks to the GConv, to be fully mapped on the FPGA for every layer on the CNN.

For MobileNetv2 (with 0.5x parameters), although our partition only considers a sequential execution of the diverse tasks in the layers, in this case, there are both an increase in energy and latency performances. This speed-up and energy efficiency factor increases with the size of the IFM, as seen in Figure 4b up to 23% and 30%, respectively.

Combining the strategies from both previous partitioning and scheduling, ShuffleNetv2 (with 0.5x parameters) benefits from a speed-up factor on both model types, with and without spatial reduction. The first section of the layer incorporates a spatial reduction block that benefits from a similar gain of parallel execution. Therefore, the gain follows the same concept as the layer from SqueezeNet, but with a $DWConv3x3$ instead of a traditional $Conv3x3$. The second section of the layer repeats a sequential execution with no spatial reduction. As a consequence, the result is similar to the layers from MobileNetv2. Because of this connection, it has the highest speed-up factor of 25% and energy efficiency of 21% compared to its homogeneous GPU counterpart as seen in Figure 4c.

Table I shows the speed-up factor and energy performance comparison with some works from Section II. This work demonstrates a similar performance, showing clear heterogeneity-related gains. Notice that the evaluated algorithms are more complex than compared state-of-the-art, achieving similar results. Therefore, because of the high parallel deployment for inference tasks, the use of FPGA-GPU heterogeneous embedded platforms also for mobile DL CNN topologies is justified, and shall result in very high gains if GPU and FPGA substrates are put closer to each other than in the tested multi-board setup.

| Work | Heterogeneous platform | | Partitioning granularity | Evaluated algorithms | Energy Gain | Latency Speedup |
|---|---|---|---|---|---|---|
| Qasaimeh, M. et al. [8] | GPU+CPU | Nvidia Jetson TX2 | Fine (Element-wise) | Background substraction | 1.74x | - |
| | | | | Color segmentation | 1.86x | - |
| | FPGA | Xilinx ZCU102 | | Harris corners tracking | 3.94x | - |
| | | | | Stereo block matching | 8.83x | - |
| Hosseinabady, M. et al. [9] | GPU+CPU | Nvidia Jetson TX1 | Fine (Element-wise) | Histogram | 1.45x-2.29x | 1.18x-1.79x |
| | FPGA+CPU | Virtex-7 and Xilinx Zynq Ultrascale+ MPSoC | | Dense Matrix-Vector multiplication | 0.96x-1.19x | 1.22x-1.48x |
| | | | | Sparse Matrix-Vector multiplication | 1.1x-1.23x | 1.15x-1.25x |
| Yuexuan Tu, et al. [10] | CPU+GPU | Nvidia Jetson TX2 | Coarse (Feature extraction +Classification) | CNN (N=16) | 2.11x | 1.3x |
| | FPGA | Xilinx Nexys Artix 7 | | CNN (N=32) | 1.94x | 1.19x |
| | | | | CNN (N=64) | 1.9x | 1.17x |
| This work | CPU+GPU | Nvidia Jetson TX2 | Mild (Layer-wise) | SqueezeNet's Fire | 1.34x | 1.01x |
| | FPGA | Intel Cylone 10 GX | | MobileNet's v2 Bottleneck | 1.55x | 1.26x |
| | | | | ShuffleNet's v2 Stage | 1.39x | 1.35x |

TABLE I: Energy and latency comparison with state-of-the-art partitioning techniques on heterogeneous FPGA-GPU implementations.

## VI. CONCLUSIONS

In this work, we have proposed, experimented and evaluated partitioning and scheduling of pre-trained mobile CNN architectures on an FPGA-GPU embedded heterogeneous platform. We have demonstrated that an FPGA exploiting Direct Hardware Mapping (DHM) outperforms a GPU implementation on a small piece of network at the cost of high resource requirements. We have also shown that the considered DL workloads benefit from a heterogeneous FPGA-GPU infrastructure when partitioned at a layer-level granularity. Indeed, the designed heterogeneous systems all outperform a homogeneous GPU-only solution either in energy, latency or both on inference for classification tasks. These results call for new fully programmable architectural solutions for embedded deep learning combining reconfigurable logic and streaming multiprocessor architectures.

ACKNOWLEDGMENT

This project has received funding from the European Union's Horizon 2020 research and innovation program under the Marie Skłodowska-Curie grant agreement No 765866